\documentstyle[12pt,epsfig]{article}
\newcommand{\be}{\begin{equation}}
\newcommand{\ee}{\end{equation}}
\def\15{\raise.9ex\hbox{1-5)\kern-0.47em}\kern.4em}
\def\69{\raise.9ex\hbox{6-9)\kern-0.47em}\kern.4em}

\begin{document}
\begin{center} 
{\LARGE Performance of a large limited streamer tube cell in drift mode}

\vspace {1cm}
{G. Battistoni}$^{1}$,
{M. Caccia}$^{1}$,
{R. Campagnolo}$^{1}$,
{C. Meroni}$^{1}$,
{E. Scapparone}$^{2}$.

\vspace {6pt}
{\it

$^1$Dip. Fisica dell' Universit\`a di Milano and INFN Sez. di Milano - Italy\\
$^2$INFN, Laboratori Nazionali del Gran Sasso, Assergi - Italy }

\baselineskip=14.5pt
\vspace {3cm}
\begin{abstract}
The performance of a large (3x3 $cm^2$) streamer tube cell in drift mode is 
shown. 
The detector space resolution has been studied using cosmic muons crossing
an high precision silicon telescope. The experimental results are compared
with a GARFIELD simulation. 
\end{abstract}
\end{center}

\baselineskip=17pt

\vspace{3.3cm}

\newpage

\section{Introduction}

In view of the completion of the first experimental phase at the Gran Sasso
underground laboratory, new
detector requirements are emerging, in the framework of possible activities
with atmospheric neutrino detectors, long baseline neutrino beam from CERN\cite{ngs} (CNGS), or 
(at present in a more speculative way) at neutrino factories.
In all the cases, although for different reasons,
muon spectrometers,  identifying
muon charge, and possibly to provide some momentum reconstruction\cite{opera,
icanoe} are required. 
The small neutrino cross section reflects in large area and/or volume
experiments: it is therefore mandatory to rely
upon detector choices which can assure reasonable accuracies on
the basis of a well assessed technology, reducing as much as possible
the number of electronic channels to be implemented.
The experience acquired with the already existing underground detectors
suggests possible solutions. 
For example, 
streamer tube chambers\cite{tube} with a cell size of
about 3x3 $cm^2$, similar to those developed for the MACRO
experiment\cite{macro}, could be operated as 
drift tubes. These devices were originally operated in digital mode,
with wire and pick up strip readout in two coordinates, for the recognition of 
charged track pattern, with a point resolution of about 1~cm. Therefore
they 
have been operated with simple electronics and without stringent  stability 
and uniformity monitoring. 
Furthermore,
these devices have been exhibiting a remarkable reliability,
being capable of continuous operation for more than ten
years. 

The space resolution required by
a massive muon spectrometer analyzing muons from neutrino interaction in a 
Long Base line experiment
is limited by the multiple scattering occurring in the magnetized iron:
a space resolution of $\sim$ 1mm is considered satisfactory.
Similar considerations apply to the case of atmospheric neutrino 
detectors.
For this reason it is worthwhile to investigate the attainable space
resolution of these devices.
It is already known in literature how the streamer regime
can be advantageous for drift operation, provided that a low rate
environment exists\cite{strdrift}: in that paper the results obtained 
using a streamer tube cell with cross section (0.9x0.9)~$cm^2$ were reported,
showing a resolution $\sigma_{x}$$\simeq$150$\mu$m.
The goal of this paper is to show that large size cell streamer tubes,
constructed in the ``coverless'' field configuration\cite{tube},
even if their electric field cannot be
considered as optimal and even if the accuracy in their
mechanical construction is not better than 0.1 mm, can achieve 
the required space resolution. 

According to the present safety rules, it is not conceivable to 
reconvert the already existing tubes of MACRO since they are built
in PVC. However, they can be constructed with a different, non dangerous,
thermo--plastic material, for instance ABS(Acrylonitrile-Butadiene-Styrene), 
always ensuring fabrication costs which are advantageous with respect to other wire-device solutions.

To measure the resolution attainable in drift mode with a streamer
tube device with large cells, we made a specific test on
an 8-wire ``coverless tube'' chamber of the same kind as used in the 
MACRO experiment, with the
only difference of being 50 cm long.
The test has been performed selecting cosmic ray tracks
by means of a high precision telescope realized with silicon pixel
arrays.
After the description of the test set-up, we shall describe the calibration of
the measurement and the experimental results.

\section{Set up description}

\begin{figure}[ht]
 \begin{center}
  \mbox{\epsfig{file=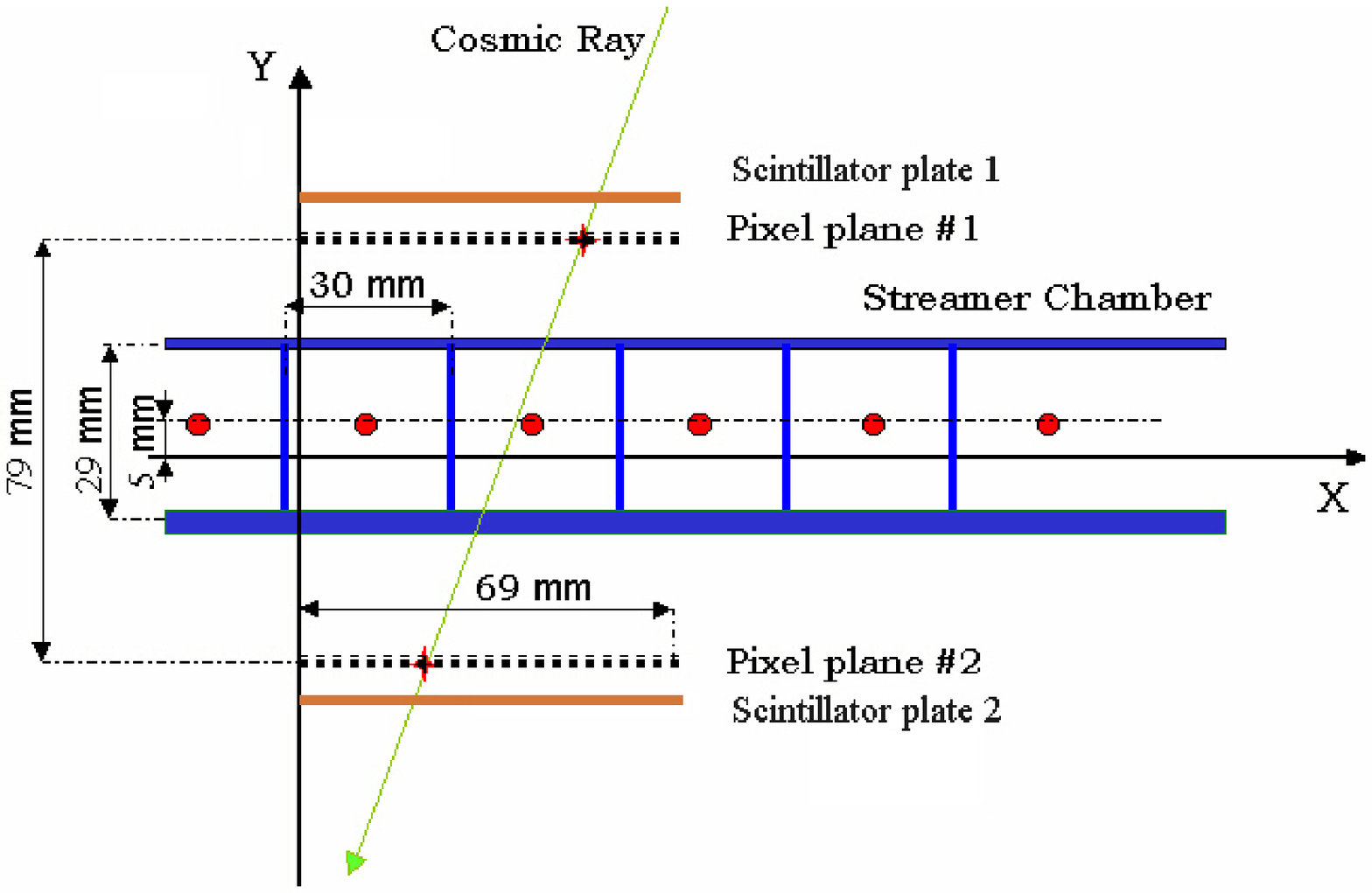,width=14cm}}
  \caption{\em Sketch of the setup used for the measurement.
\label{fig:setup}}
  \vspace{-0.5cm}
 \end{center}
\end{figure}

The experimental setup for the test consisted of a high precision cosmic ray 
telescope and the streamer chamber under
test  (Fig.\ref{fig:setup}). 
The telescope was built using two planes of silicon pixel arrays and two 
plastic scintillators (20x70x10) mm$^3$ for triggering purposes.
The silicon pixel detectors used  were arrays of 8064 pixels with dimensions 
of (330 x 330) $\mu$m$^2$, and a thickness of about 300$\mu$m. 
These pixels are readout in digital mode, providing thus
only the coordinates of the pixels hit by a particle. 
These detectors are the same as the ones used 
in the DELPHI Very Forward Tracker 
(VFT, \cite{siltrack,becks}).  
The active area of 990 mm$^2$ is divided into 
10 regions of 24x24 pixels, followed 
by six regions of 24x16 pixels, for a total length of 69mm and a width of 
22mm and 17mm respectively. 
Typical figure of merit of these detectors are $\epsilon\approx~99\%$, 
with a threshold around 10,000 electrons, a number of noisy pixels at 0.3\% 
level and a point resolution of 100$\mu$m in both coordinates.

\begin{figure}[ht]
 \begin{center}
  \mbox{\epsfig{file=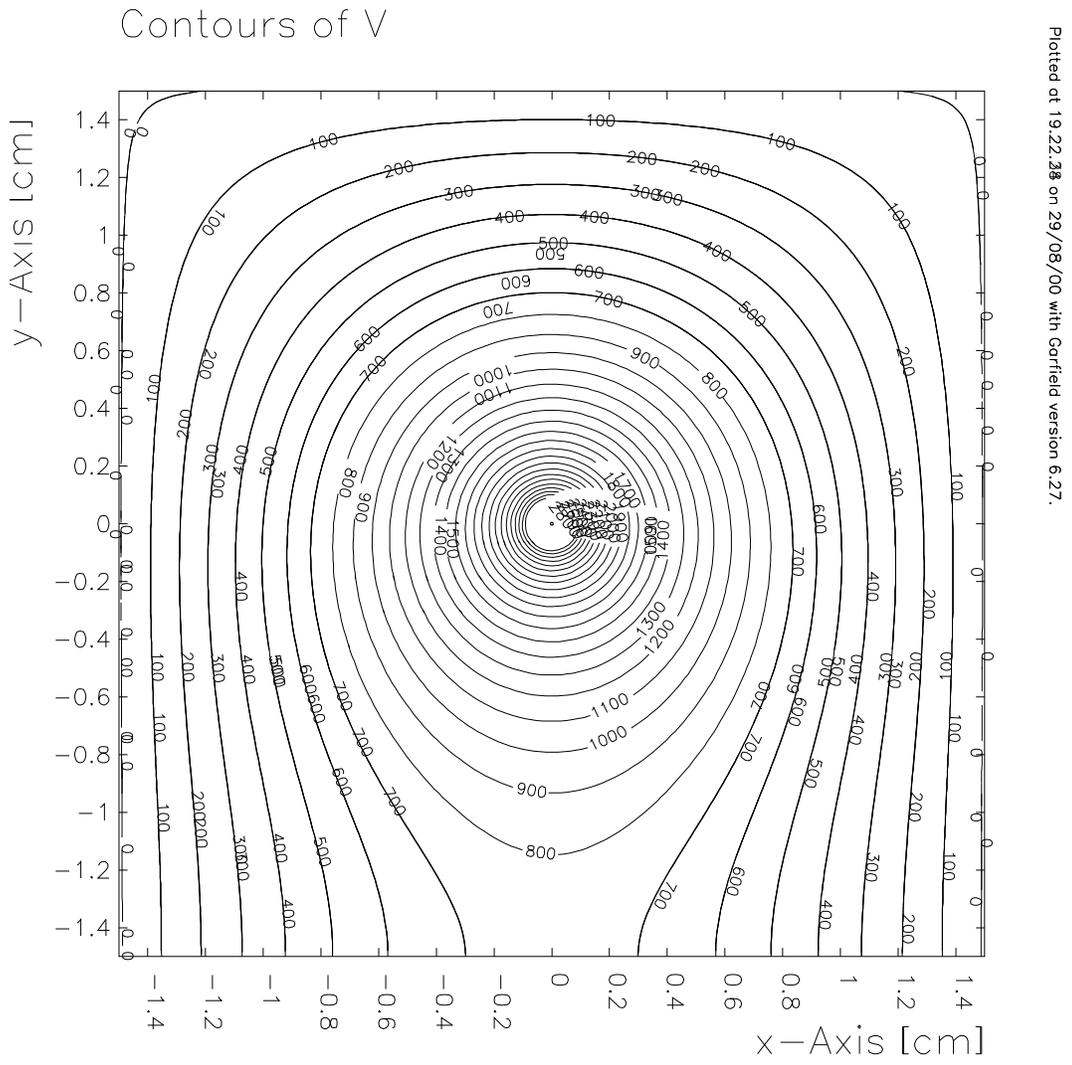,width=14cm}}
  \caption{\em  Equipotential line in the streamer tube cell
operating at the 5500 V anode voltage.   
  \label{fig:garfield1}}
  \vspace{-0.5cm}
 \end{center}
\end{figure}

A 50 cm long chamber  of the same kind of those used in MACRO has been 
positioned 
in the middle plane of the telescope (Fig. \ref{fig:setup}).
The  description  of the  plastic  streamer  tubes
of  MACRO has been extensively  reported in \cite{macro} 
and references therein.  We remind here that  the basic unit of the tube
system is a  chamber containing eight
individual cells.  The single cell is  (2.9~x~2.7)~cm$^2$.
The  anode wire
(silvered Be-Cu) has a diameter of  100~$\mu$m.
Three sides of the cell 
are coated with a low--resistivity  graphite ($<1~k\Omega$/square) to
perform the  cathode  function by the  electrode-less  shaping principle
\cite{electrodeless}. This  structure is inserted inside an uncoated PVC
envelope (1.5 mm thick) and closed  by two plastic end caps at the ends.
The overall external size of our chamber is  3.2  cm x 25 cm x 50 cm.
Each wire behaves as a transmission line with a characteristic impedance
of 330 $\Omega$ and a propagation time sligthly larger than 3.3 ns/m. 
They provide both
digital readout for  tracking  and analog  (charge) readout.
The chamber is fluxed with an Argon + Isobutane with relative fractions
of 50/50 gas mixture, 
controlled within a 15\% accuracy.
In these operation conditions, full efficiency is achieved for an anode
high voltage value of about
5500 V, with a single counting rate plateau spanning from 5200 V to 5900 V.
The operation properties of streamer tubes operating with various gas
mixtures has 
been extensively studied in \cite{tube,macro,iaroccivari} and 
references therein. 
The field configuration of the coverless MACRO tube cell 
can be derived from
Fig. \ref{fig:garfield1} which shows the equipotential 
lines as obtained with the GARFIELD program\cite{garfield} for an anode
voltage of 5500 V.

The geometry of the setup is sketched in Fig. \ref{fig:setup}.
A cartesian coordinate system is defined , having the x axis
parallel to the long side of the pixel detectors,
the z axis parallel to the short side of the pixel detectors
and to the wires of streamer tube. As a consequence the y axis 
is perpendicular to the plane of the detectors.
Since the pixel detectors are only 69 mm long, the trigger area 
covers 2  cells and the edge region of the adjacent ones.
All the measurements reported here have been performed with a
distance between the silicon planes of 79 mm. In this configuration,
we triggered only tracks with an angle with respect to the vertical not
exceeding 35$^\circ$, with a most probable value around $10^\circ$.
The cosmic trigger rate was of $\sim$1 track/min. 

The accuracy in the determination of the track position at the wire level, 
taking into account the
multiple scattering in the silicon pixels and in the streamer tube walls, 
is $<$90 $\mu$m, assuming an average muon energy of
$\sim$1 GeV.
The trigger signal is provided by the coincidence of the scintillator
pulses.
The telescope is readout by means of a Pixel Readout Repeater, 
driven by a Pixel Camac Board\cite{siltrack},
that provides the necessary timing signals and the power supply lines.
The signal from the tube wires are OR-ed together by means of a 
resistive network, providing an output signal on 50 $\Omega$ impedance.
The wire pulses are properly discriminated ($50 mV/50 \Omega$) and sent
to a TDC (Lecroy 2228A), which records the
time difference between the scintillator trigger signal and the
streamer chamber signals. The TDC was used with a 1bit resolution 
of 250ps.

The acquisition chain has been realized by means of a Labview 
application, developed in our group, 
and was controlled by a PC equipped with a GPIB CAMAC 
controller (Lecroy 8901 A). 
The acquisition included also
a scaler (Lecroy 2551) to record the counts of the scintillators 
and of the singles tubes, in order to perform stability checks.
The Pixel Readout runs continuously a gate--reset sequence
until a scintillator trigger in coincidence with the gate is received.
In this case the first bit of an input/output register is used
to control the synchronization between the PC and the Pixel Camac Board.
The acquisition monitoring is provided by a graphical interface.

\section{Detector calibration}
\begin{figure}[ht]
 \begin{center}
  \mbox{\epsfig{file=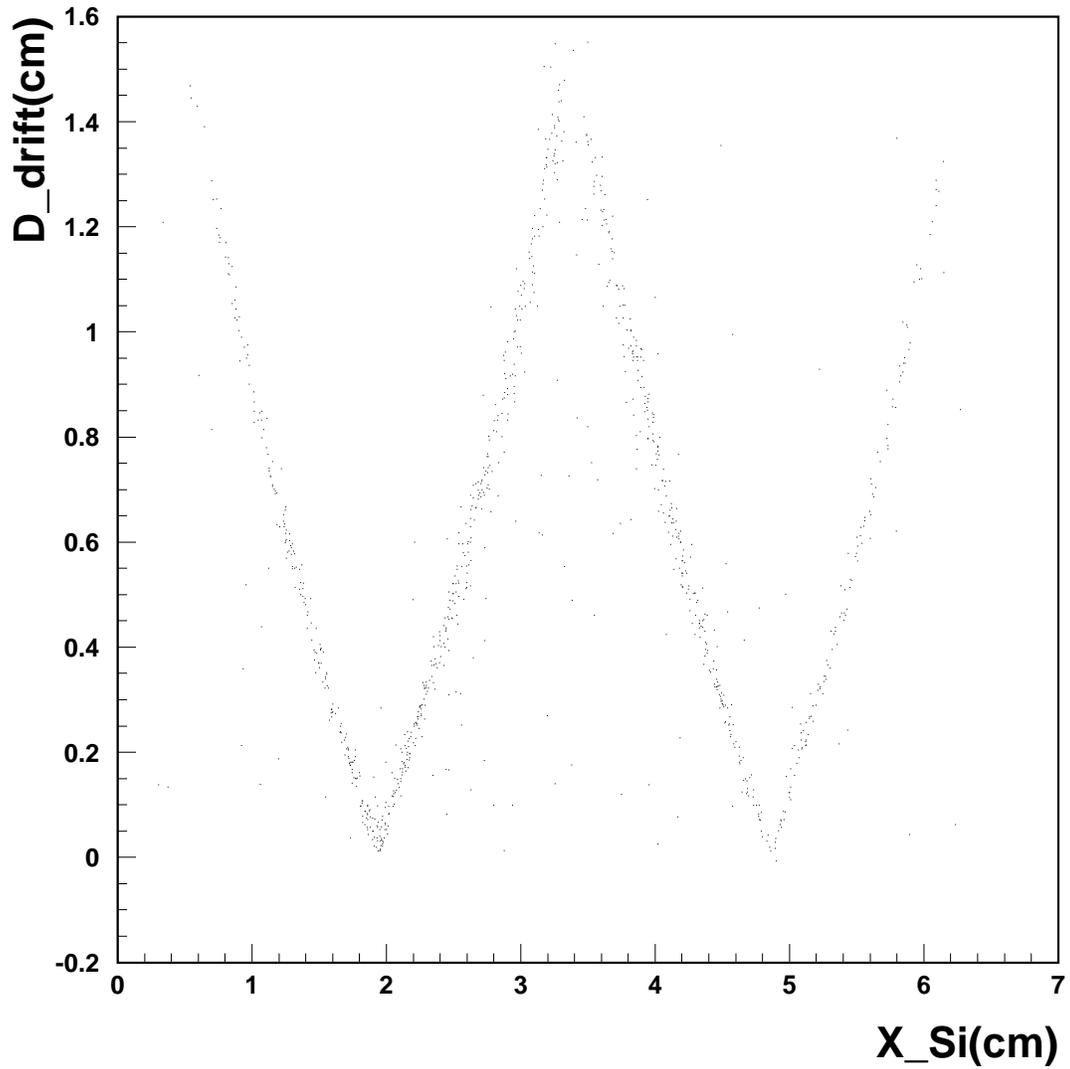,width=14cm}}
\caption{\em Track distance from the wire, as reconstructed with
the time measurement, as a function of the position in the cell
as reconstructed from the pixel telescope.
\label{fig:figuraV}}
  \vspace{-0.5cm}
 \end{center}
\end{figure}

As a first step, we used a small data sample to have a precise geometrical
survey of  the actual position of the wires. Fig.~\ref{fig:figuraV} shows the
drift distance as a function of the X position computed at the wires level Y=0, 
measured by using the silicon pixel detector. 
The wires are of course positioned where the drift distance 
reaches his minimum, {\it i.e.}
at the lower cusp points of the v--shaped distributions.
By means of a linear fit to the four straight portions of 
this scatter plot we obtained
$X_{wire~1}$~=(1.95$\pm$0.05)cm  and $X_{wire~2}$~ =(4.85$\pm$0.05)cm. 
A first measurement of the average drift velocity $V_{drift}$ 
(neglecting differences as a function of the distance from the wire)
can be obtained by selecting vertical
tracks and imposing:
\begin {equation}
Max(TDC)\cdot R_{TDC}\cdot V_{drift} = Cell/2, 
\end{equation}
where Max(TDC) is the maximum value
of the TDC in the data sample, $R_{TDC}$ is the conversion factor from 
TDC channels to ns and Cell/2 is half of the cell length.
The actual cell width of each tube is known with an accuracy of 0.1 mm.

The value of $R_{TDC}$ has been calibrated to be
$R_{TDC}$=0.242ns/channel and 
$R_{TDC}$=0.248ns/channel for the first and the second cell respectively.
The resulting average drift velocity from the above mentioned data sample
was found to be $V_{drift}$=(4.5 $\pm$ 0.3) cm/$\mu$s.

Data were collected at different operation high voltage, 
ranging from 5500~V to 5900~V,
by using simultaneously two tube cells. The gas mixture was kept
unchanged.

We do not expect a constant drift velocity as a function
of position. Therefore, in order to derive a more accurate
estimate of the space-time relation, we adopted the following method
for each different set of operation conditions.

\begin{figure}[ht]
 \begin{center}
  \mbox{\epsfig{file=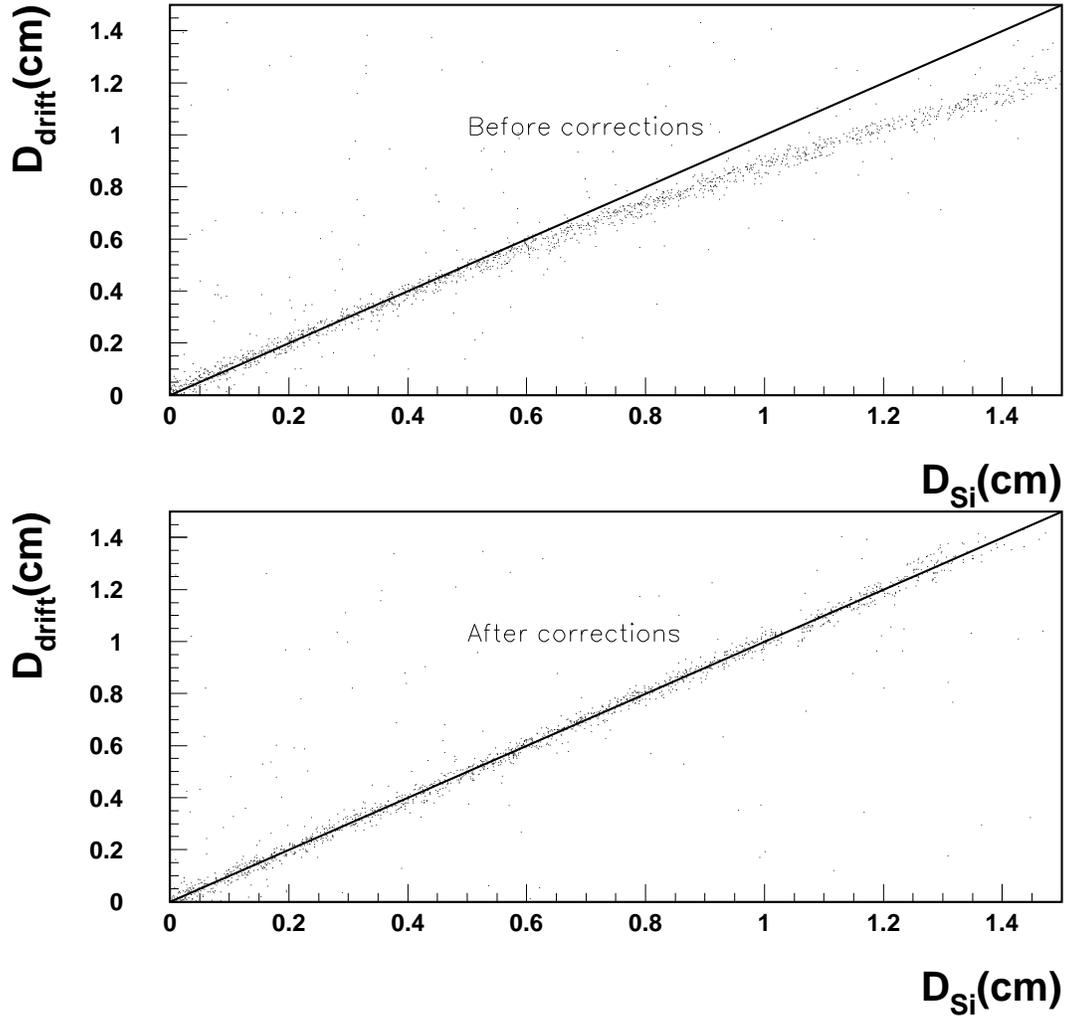,width=14cm}}
  \caption{\em Distance from the wire as obtained from the
  drift measurement vs the distances obtained from the pixel telescope.
Upper panel: results using a drift velocity averaged on the cell
  width. Lower panel: after corrections applied to take into account the
  variation of drift velocity as a function of the distance from the wire.
The straight line is put to guide the eye.
  \label{fig:correzione}}
  \vspace{-0.5cm}
 \end{center}
\end{figure}

We have computed the minimum track-wire distance in two ways:
\begin{enumerate}
\item $D_{drift}$, using the
drift time measurement (according the average drift velocity quoted
above); 
\item $D_{Si}$, using the track reconstructed from pixel readout.
\end{enumerate}
If we plot $D_{drift}$ versus $D_{Si}$ as in the top panel
of Fig. \ref{fig:correzione},
a clear deviation from linearity is observed.
We made the assumption that the required correction depends only on the 
cell geometry, the gas mixture and the high voltage, being independent of the
specific cell used. Under this hypothesis we  considered together the data
from the two cells and a correction as a a function of $D_{Si}$ was
introduced to restore linearity.
Such correction is achieved by fitting the scatter plot
with a piece-wise linear fit, so that
 2 parameters for each $D_{Si}$ section are obtained from the law:
\begin{equation}
L(cm) = p_0 + p_1\cdot TDC\cdot R_{TDC}\cdot V_{drift} 
\end{equation}

The  $p_0$ and $p_1$ values are reported in Table 1.
\begin{table*} [hbt]
\begin{center}
\begin {tabular} {|l|l|l|l|}
\hline Distance from wire (cm)  &$p_O$ (cm)    &$p_1$   \\
\hline
 0.   -- 0.65   & -0.014   & 1.13   \\
\hline                           
 0.65 -- 1.0    &  0.266   & 1.02   \\
\hline
 1.0  -- 1.5    &  0.016   & 0.95    \\
\hline                           
\end{tabular}
\caption{\it Values of fitted parameters in the correction algorithm
for the three different distance regions at the 5500 V operation point.}
\label{tab:magne}
\end{center}
\end{table*}

The bottom panel of Fig. \ref{fig:correzione}
shows the distribution of data after correction.
Such procedure can be applied to any streamer tube cell working in an
actual
experiment. 

\section{Experimental results}

The overall space accuracy from the drift time measurement
can be obtained from the 
distribution of the residuals of the fit of Fig. \ref{fig:correzione}
(bottom panel), summing up all cells.

\begin{figure}[ht]
 \begin{center}
  \mbox{\epsfig{file=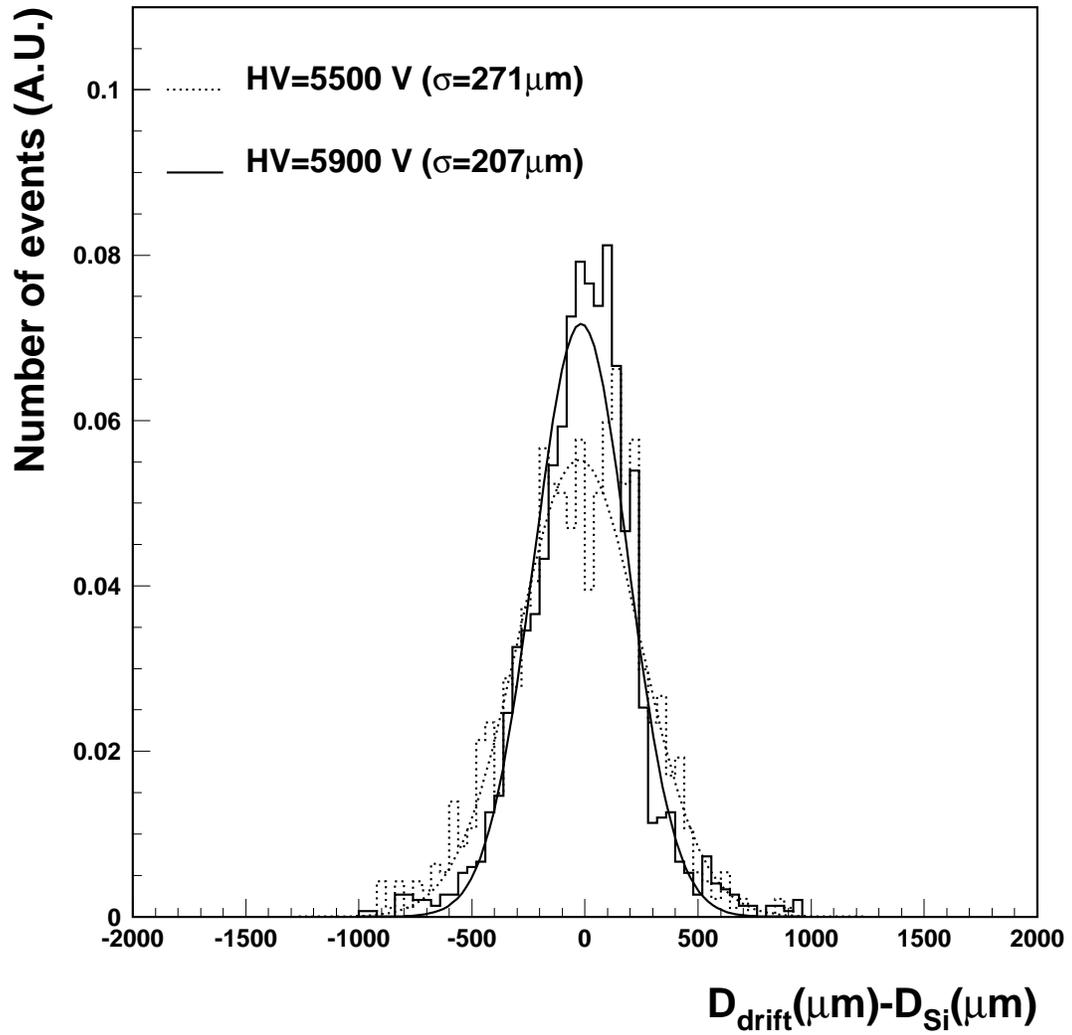,width=14cm}}
  \caption{\em Distribution of the difference between the track position 
obtained from the drift measurement and from the position detector.
  \label{fig:sigma}}
  \vspace{-0.5cm}
 \end{center}
\end{figure}

The results are shown in Fig. \ref{fig:sigma} for HV~=~5500~V and 5900~V.
The gaussian fits give a resolution of
$\sigma_{tot}$=(271$\pm$7)$\mu$m and 
$\sigma_{tot}$=(207$\pm$5)$\mu$m for HV~=~5500,~5900 V respectively.
The standard deviation of such distributions can be
written as $\sigma_{tot}$=
$\sqrt {   \sigma_{drift}^2 +  \sigma_{Si}^2  }$=
271 $\mu$m (207 $\mu$m).
Since the silicon pixel size is 330 $\mu$m the ideal 
contribution of 
$\sigma_{Si}$ to $\sigma_{tot}$ would be
$\sigma_{Si}$=330~$\mu$m/$\sqrt{12}$/$\sqrt{2}$=68~$\mu$m and 
hence the intrinsic space resolution achievable with the
drift time method would be $\sigma_{drift}$=
$\sqrt{\sigma_{tot}^2-\sigma_{Si}^2}$=262 $\mu$m~(195 $\mu$m). 
Taking into account also the expected muon momentum distribution 
($<E_\mu>\sim$~1~GeV) and 
the multiple scattering inside the silicon and the chamber walls,
we obtain an average value of $\sigma_{Si}$=94. The estimated value of 
$\sigma_{drift}$ then improves and it is $\sigma_{drift}$=254 $\mu$m 
(184 $\mu$m).

\begin{figure}[ht]
 \begin{center}
  \mbox{\epsfig{file=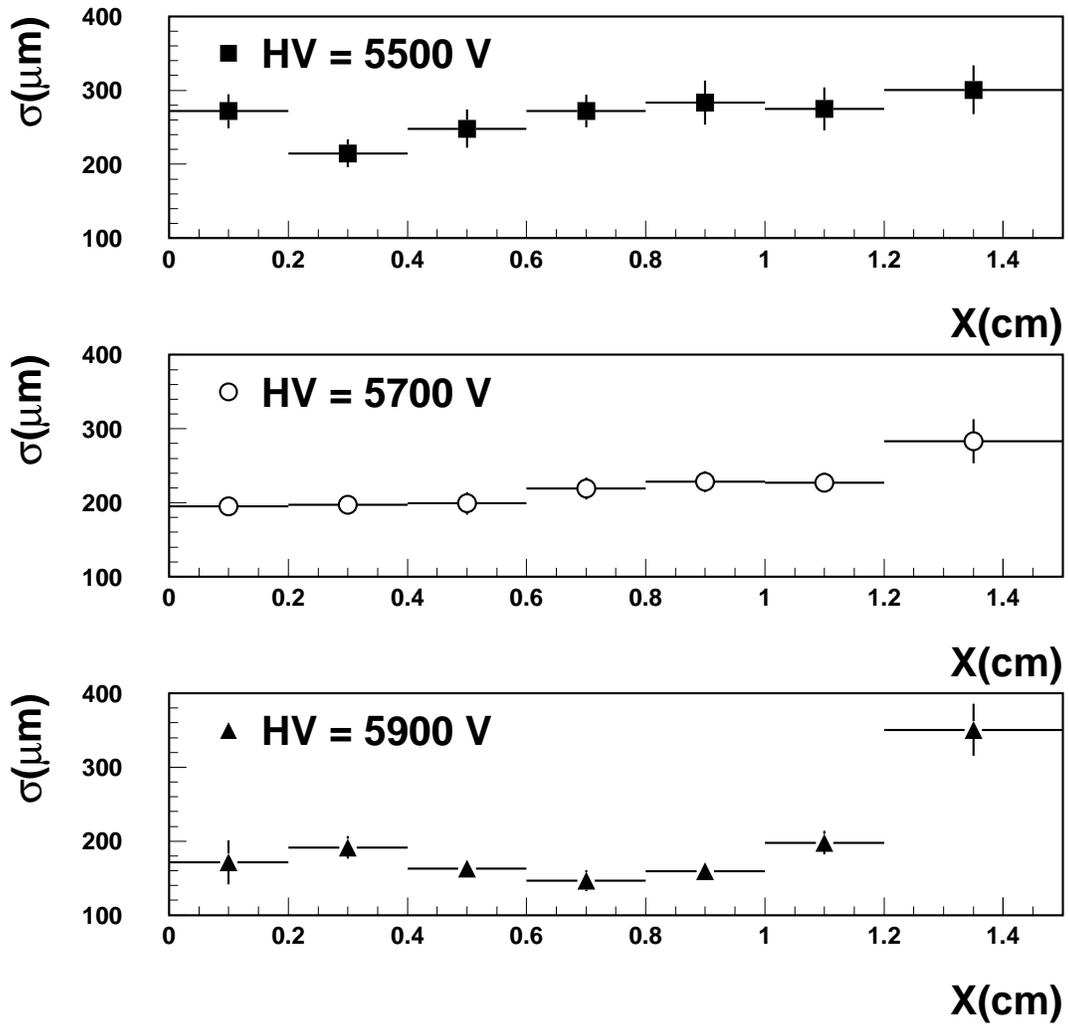,width=14cm}}
  \caption{\em Position resolution achieved with the streamer tube in drift
  mode as a function of the coordinate along the wire plane, for three
  different operation voltages.
  \label{fig:tresigma}}
  \vspace{-0.5cm}
 \end{center}
\end{figure}

We also tried to measure
$\sigma_{drift}$ as a function of the X distance from the wire, 
selecting different track samples on the basis  of the track parameters
reconstructed with the silicon detector. 
In order to maintain
an high efficiency and reliability of the streamer tubes system in
large area detectors, the HV has to be kept within the plateau, so we did
not attempt to go beyond HV=5900~V. 
The results are shown in Fig. \ref{fig:tresigma}.
We can identify at 5900 V an optimal $\sigma_{drift}$ 
in the central part of the tube, i.e. X$\simeq$ 0.7 cm. 
At lower X, 
the resolution is spoiled by a probable non linear space--time relation,
while at large X it is affected by the 
reduced electrical field in proximity of the cell walls and corners.
The resolution improves increasing the HV, due to the reinforcement
of the electric field in the cell wall neighbourhood. 
 
\begin{figure}[ht]
\begin{center}
\mbox{\epsfig{file=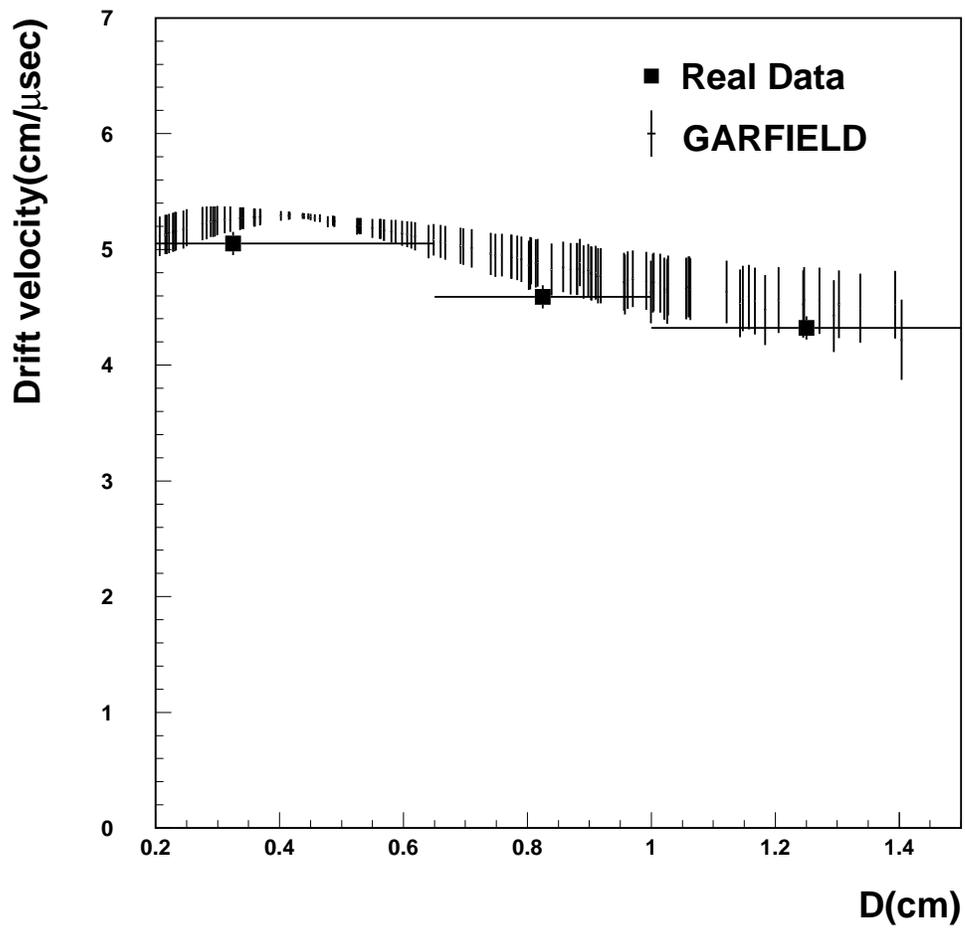,width=14cm}}
\caption{\em Comparison of the expected drift velocity (GARFIELD) with 
the present results.    
\label{fig:garfield2}}
\vspace{-0.5cm}
\end{center}
\end{figure}

In order to verify these hypotheses,
we simulated the expected drift velocity as a function of distance from
the wire using the GARFIELD package.
To take into account the trigger acceptance, we used
an angular distribution taken from real data tracks, reconstructed with the
silicon pixel detector. The (x,y) position corresponding to the minimal 
distance from the wire 
was estimated for each track and used as input to the GARFIELD simulation.  
The curve in 
Fig.\ref{fig:garfield2} shows the simulated drift velocity  as a function
of the distance from the wire, for HV=5500 V. The simulation 
is obtained changing  the Isobutane fraction in the gas mixture from 
40$\%$ to 60 $\%$. 
An higher fraction of quenching gas reflects in a smaller drift velocity
at low electric field (i.e. near the tube walls).
The experimental data are in satisfactory 
agreement with the GARFIELD simulation.
The space resolution quoted above could be in principle slightly spoiled in a 
longer streamer cell, due to the wire sagitta.

\section{Discussion and conclusions}

The performance of a large streamer tube cell in drift mode have been
tested using an high precision silicon telescope: space resolution
between 150 $\mu$m and 350 $\mu$m have been obtained, depending on 
the HV applied and on the track distance from the wire.

Such quite good resolution, can be anyway spoiled,
when using streamer tubes in drift mode for real spectrometers:\\
a)  In case of longer tubes, the practically 
achievable
resolution is deteriorated by the wire sag position determination. 
This problem is always present in any long tube: the relevance of such
effects depends on the geometry of the tube. Should the magnetic field 
deflection lye in the same plane where the gravitational force applies,
the resolution deterioration would be more severe.\\
b) Concerning non superconducting spectrometer, equipped 
with iron, the space resolution
is deteriorated by $\delta$-rays and electrons-positrons pairs produced in 
the magnetic field iron yoke, penetrating inside the cell. 
When a muon crosses the cell close to the walls, 
such particles may result in earlier streamer
avalanches, producing thus shorter drift times.\\ 
c) As it has been pointed out in \cite{strdrift}, inclined tracks may
hit streamer tube cells at different coordinates along the wires, 
introducing thus additional delays. 
Due to the relatively fast signal velocity transmission along
the wire, $\simeq$25 cm/ns, compared to standard drift velocity, a rough 
knowledge of the coordinate of the 
event along the wire, ($\simeq$ 1m) is required to recover the full 
resolution.  

Summarizing, 
although realistic spectrometers equipped with
streamer tubes operating in drift mode, may suffer the space resolution 
deterioration quoted above, the expected resolution is
anyway well below 1 mm. Taking into the account the multiple scattering
suffered by muons in the iron, such resolution is still adequate 
to fulfill the requirements for massive 
iron spectrometers for neutrino physics, indicating streamer tubes
in drift mode as a possible choice for particle tracking in 
muon charge and momentum analysis.

\end{document}